\documentclass[fleqn,10pt]{wlscirep}
\usepackage[utf8]{inputenc}
\usepackage[T1]{fontenc}
\usepackage{amsmath,amssymb,amsfonts}
\usepackage{bm}
\usepackage{graphicx,}
\usepackage{verbatim}
\usepackage{float}
\usepackage[FIGTOPCAP]{subfigure} 
\usepackage{dcolumn}
\usepackage{hyperref}
\usepackage{tikz}
\usetikzlibrary {arrows.meta} 
\usetikzlibrary{patterns}
\usetikzlibrary{hobby}
\usepackage{appendix}
\usepackage{ulem}

\newcommand{\RN}[1]{%
	\textup{\uppercase\expandafter{\romannumeral#1}}%
}

\title{Single replica spin-glass phase detection using field variation and machine learning}

\author[1]{Ali Talebi}
\author[1]{Mahsa Bagherikalhor}
\author[1]{Behrouz Askari}
\author[1,2,*]{G.Reza Jafari}
\affil[1]{Department of Physics, Shahid Beheshti University, Evin, Tehran 1983969411, Iran}
\affil[2]{Center for Communications Technology, London Metropolitan University, London N7 8DB, UK}

\affil[*]{gjafari@gmail.com}

\begin{abstract}
The Sherrington-Kirkpatrick spin-glass model used the replica symmetry method to find the phase transition of the system. In 1979-1980, Parisi proposed a solution based on replica symmetry breaking (RSB), which allowed him to identify the underlying phases of complex systems such as spin-glasses. Regardless of the method used for detection, the intrinsic phase of a system exists whether or not replicas are considered. We introduce a single replica method of spin-glass phase detection using the field's variation experienced by each spin in a system configuration. This method focuses on a single replica with quenched random couplings. Each spin inevitably observes a different field from the others. Our results show that the mean and variance of fields named "Spontaneous Configurational Field" experienced by spins are suitable indicators to explore different ferromagnetic, paramagnetic, and mixed phases. To classify different phases of the system with defined indicators we have developed an algorithm based on machine learning to analyze the desired samples.
\end{abstract}
\begin{document}

\flushbottom
\maketitle
% * <john.hammersley@gmail.com> 2015-02-09T12:07:31.197Z:
%
%  Click the title above to edit the author information and abstract
%
\thispagestyle{empty}

\section*{Introduction}
%paragraph 1 
The spin-glass, a disordered magnetic system, contains ferro- and antiferromagnetic couplings between each pair of spins that cause frustration in updating spins to reach a stable state. Edwards-Anderson model was the first model for describing the behavior of spin-glass systems~\cite{AndersonEdwards}, The model includes the interaction of a spin with its nearest neighbors. In 1975, an exactly solvable model with long-range interaction was introduced by Sherrington and Kirkpatrick(SK) as a mean field simplification of the version of the Edwards-Anderson model~\cite{SK1, Sherri}. Spin-glass systems are complex and challenging to study due to their disordered nature and the frustration in their interactions. Researchers have developed various methods to understand these systems better. Some of the approaches to deal with the intricacies of spin-glass systems are the Replica method~\cite{ParisiReplica, PARISI1979, VanHemmen}, Replica Symmetry Breaking (RSB)~\cite{Parisi1, Binder1986}, TAP equation~\cite{TAP}, and the Cavity method~\cite{cavity}. In the Replica method, the free energy calculation is simplified by considering multiple system replicas and averaging them. It assumes that replicas are independent and identical. However, the trick works well at high temperatures, faces errors in low temperatures, and fails to predict the nature of the spin-glass phase~\cite{Spin_Glass_Theory_and_Beyond, Nishimori, Parisi1}. Later, Parisi introduced a replica symmetry breaking(RSB) solution to address the limitations of the replica method\cite{Parisi4, Parisi3, Parisi1980, Parisi1}. Replicas are not identical anymore, and this provides a more accurate description of the spin-glass behavior. This method can successfully reveal many well-separated local minima of the system, which corresponds to different systems's configurations and plays a crucial role in understanding the phases of the disordered complex systems. 

Studying spin-glass systems to understand their complex behavior has been challenging for researchers over the years, and its application has led to significant advancements across various scientific disciplines, including condensed matter physics~\cite{PhysRevLettTam, Drozd,park2022observation}, materials science~\cite{Mydosh2015,katukuri2015strong}, neuroscience~\cite{Hudetz2014,knuuttila2014}, and quantum computing~\cite{Lidar, Callison2019,kadowaki1998quantum,king2023quantum,harris2018phase,knysh2016zero,grass2016quantum}. However, the SK model has continuous phase transition and typically considers pairwise interactions between spins; in many real-world systems, interactions can involve more than two spins, leading to higher-order terms in the Hamiltonian~\cite{GROSS, GARDNER, Crisanti, Mahsa2}. The quantities that can help to identify different phases of the system are order parameters. In spin-glass systems, magnetization and overlap can determine the system's behavior, where magnetization defines a measure of the average magnetic moment per spin, and the overlap measures the similarity between different spin configurations. In the SK model of the spin-glass system, to shed light on the system's behavior, the mean-field technique was introduced by Thouless, Anderson, and Palmer~\cite{TAP}, which is called the TAP equation. Another method known as the Cavity method was developed by Mézard, Parisi, and Virasoro, in which the average field applied to one spin of the network from other spins of the network was related to the average field applied to the same spin if it removed from the network~\cite{cavity}.

\begin{figure}
	\centering
	\includegraphics[scale = 0.55]{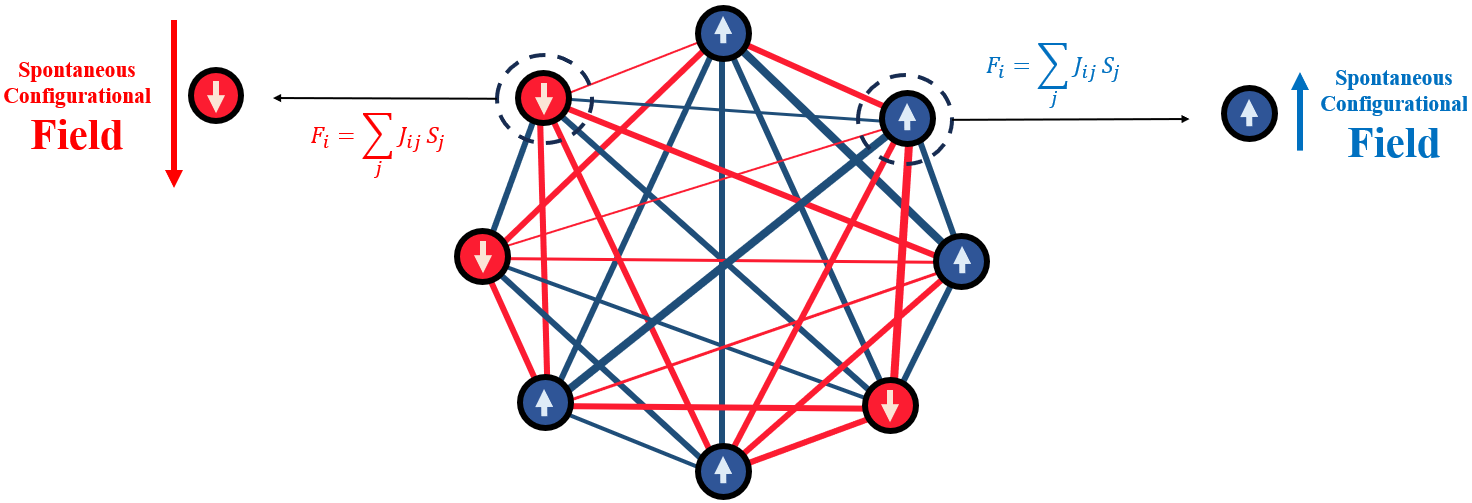}
	\caption{The image shows a system's configuration, with blue circles representing $+1$ spins and red circles symbolizing $-1$ spins. Furthermore, the positive and negative interactions are depicted by blue and red lines. The arrows indicate spontaneous configurational fields based on the Eq.(\ref{eq3}) with different lengths, illustrating the direction and intensity of fields.}
	\label{fig:fig1}
\end{figure}

All these approaches are used to unravel the intricate behavior of spin-glass systems and help us understand their different phases. Based on the literature, we identify three key parameters: temperature, mean, and variance, which refer to the mean and the variance of the Gaussian distribution function in which random interactions, $J_{ij}$, are taken. However, for phase classification, only two ratios, mean to variance and temperature to variance, are essential independent variables. Detecting different phases of spin-glass systems can be done in various approaches such as theoretical, numerical, experimental, or a combination of these methods. In this study, we use a simulation-based approach.
While multiple replicas are typically needed to explore phases, our study focuses on a single replica with quenched random couplings. The model presented here focuses on the field felt by each spin from others via random interactions in a system's snapshot Fig.(~\ref{fig:fig1}). Indeed, in the context of spin-glass phenomena, there is a notable degree of disorder in the interaction between the spins, which gives rise to a discrepancy in the field observed by each spin. It will be demonstrated that this discrepancy, along with the field, which in this article we named Spontaneous Configurational Field, felt by each spin, plays a pivotal role in determining the overall separation of the phases observed in spin-glass, ferromagnetism, paramagnetism, and mixed phase. Finally, with machine learning, a program will be devised to facilitate the identification of the phase present in unknown samples. Providing a configuration of an unknown system, we identify distinct phases by applying the devised algorithm. Additionally, this method helped us find the similarity of each point to other well-known phases, which helps find the similarity of different parts of the Mixed phase to the other three phases. 
\subsection*{SK Model}
In this section, we review the approach proposed by researchers to illuminate the complexities of spin-glass and to deepen our understanding of their phase transitions. Let's consider the Hamiltonian of the SK model in the absence of an external field.
\begin{equation}\label{eq1}
H = - \sum_{i<j} J_{ij} \ S_{i} \ S_{j},
\end{equation}
where $S_{i}$ and $S_{j}$ are spins which can take the values $\{\pm 1\}$ and $J_{ij}$ is the interaction between each pair of spins and comes from a Gaussian Probability distribution with mean $J_0/N$ and variance $J/N^2$ to confirm that the Hamiltonian is an extensive variable. According to the ratio of mean to variance of the Gaussian distribution function and the ratio of temperature to variance, the system is placed in one of paramagnetic, ferromagnetic, spin-glass, and mixed phases. Phase transitions in spin-glass with the Hamiltonian Eq.(~\ref{eq1}) observe when changes in temperature and the ratio of mean to variance lead to a shift from one magnetic phase to another, such as transitioning from a paramagnetic state to a spin-glass phase and from ferromagnet to spin-glass phase.

The first effort to study the phase diagram of the SK model is derived from the replica symmetric. This method simplifies the calculation of the expected value of the logarithmic partition function into a more manageable calculation of the expected value of the partition function to the power of $n$;  where n is an integer and number of copies of replicas of the system. However, it is based on the assumption that replicas are symmetric. Its solution typically includes phases characterized by different magnetic orders and transitions between these phases. The replica method solution results in two equations of state for the ferromagnetic order parameter, $m = \langle S_{i}^\alpha\rangle $, which is an indicator of the alignment of spins and the spin-glass order-parameter, $q = \langle S_{i}^\alpha S_{i}^\beta\rangle $, in which $S_{i}^\alpha$ and $S_{i}^\beta$ are the spin states at site $i$ in two different replicas $\alpha$ and $\beta$ and due to replica symmetric assumption they reduce to $m = \langle S_{i} \rangle$, $q= \langle S_{i} S_{i} \rangle$. The replica symmetry solution highlights regions of stability and instability in the system ~\cite{SK1,Sherri}. The overlap in the simulation approach is considered a dynamical parameter such as defined as $q_{\mathrm{EA}}=\lim _{t \rightarrow \infty} \lim _{N \rightarrow \infty}\left[\left\langle S_i\left(t_0\right) S_i\left(t_0+t\right)\right\rangle\right]$ ~\cite{Nishimori}, which shows multivalley structure in free energy of this system, and when the system finds the global minimum of the free energy it will freeze and the value of this parameter will be $+1$.  

%\subsubsection*{Replica symmetry breaking(RSB)}
By failure of the assumption of replica symmetry in low temperature, the theory of replica symmetry breaking considers various values for $m$ and $q$ depending on replica indices $\alpha$ and $\beta$ as $m_{\alpha}$ and $q_{\alpha\beta}$ to prevent the unphysical conclusion of the replica symmetric solution. 
Therefore, $q$ is not a single value but a distribution in the RSB approach and is a critical order parameter for detecting the spin-glass phase. 
In addition to these phases, the SK model features 
a mixed phase in which both ferromagnetic and spin-glass orderings coexist. The boundary of this phase, referred to as the AT-line by Almeida and Thouless ~\cite{Thouless_1978}, specifies the transition between the spin-glass and ferromagnetic phases, marking the onset of replica symmetry breaking and the presence of a glassy state. The area between the AT-line and the spin-glass phase is identified as the mixed phase, characterized by the instability of the solutions derived from the SK model. Subsequently, Parisi ~\cite{Parisi4,Parisi1980,PARISI1979,Parisi3} was able to establish a distinct boundary between the spin-glass phase and the mixed phase by exploring the concept of replica symmetry breaking.

\subsection*{TAP and cavity method}
The same phenomenon in the TAP equation~\cite{TAP}, derived by Thouless-Anderson-Palmer states that The average field felt by each spin depends on the magnetization of other spins in the network, which is the ensemble average
of the sign of a spin in the network. This method is based on an effective field $F_i = \sum_{j \neq i} J_{ij} \ \langle S_{j} \rangle$ experienced by each spin due to its neighbors, which is known as mean-field theorem. The study of the Thouless-Anderson-Palmer (TAP) variational principle concerning the SK model is a mean-field technique that focuses on the self-consistent equation Eq.(~\ref{eq2}) of order-parameter.
\begin{equation}\label{eq2}
m_i=\tanh \beta\left(\sum_j J_{i j} m_j-\beta \sum_j J_{i j}^2\left(1-m_j^2\right) m_i\right)
\end{equation}
In the cavity method~\cite{cavity}, the field experienced by each spin is influenced by its neighboring interactions. The effective field within cavities (the local environment of a spin), the field that each spin encounters when it is excluded from the network depends on the site of the spin. By reviewing the proposed methods for a better understanding of the spin-glass systems, in the following section, we address our proposed approach.
\section*{Method} 
\subsection*{Single replica method}
%\textbf{Single replica method}.
While the aforementioned approaches address the complexity of the spin-glass phase and rely on replicas, either through replica symmetry or RSB, the intrinsic properties of a system exist independent of the detection methods. These methods investigate various characteristics of spin glass systems to analyze their phase transitions, and the simulation approach mostly investigates the dynamics of these systems; however, we studied the system's behavior using only a single replica in a snapshot. This approach allows us to examine the system’s configuration at a certain time to determine its phase in our proposed method. In our study, the field each spin experiences differs from that of the others due to the random interactions represented by $J_{ij}$. The variation in these experienced fields plays a vital role in our proposed method, which is constructed on the mean and the deviation of the field each spins feels in disordered systems. Our results show that these parameters serve as excellent indicators for accurately identifying the various phases of the system. According to Hamiltonian~\ref{eq1}, each spin is experiencing a field due to its interactions with others. The field that is applied to the spin $S_{i}$ by the other spins can be written as
\begin{equation}\label{eq3}
	F_i = \sum_{j \neq i} J_{ij} \ S_{j}, 
\end{equation}
The summation over $J_{ij}S_{j}$ yields different values for each selected spin, as the interactions are random values that are taken from Gaussian probability distribution. We defined a specific field, which is calculated in a single configuration of the system capturing a 'snapshot' of the system at a specific time, when it is located in a global or local minimum, without adopting a dynamical approach, and it may be better to recognize this method as a morphological method. We named this field the Spontaneous Configurational Field (SCF) Eq.(\ref{eq3}). The mean and the variance of the fields, experienced by each spin in the system with size $N$, can be calculated as
\begin{equation}\label{eq4}
\langle F\rangle=\left\{\begin{array}{cl}
\frac{\sum_{i \in N} F_i}{N}, & \text { if } J_0/ J \leqslant 1 \\
\frac{\sum_{i \in N} F_i}{N J_0}, & \text { if } J_0/ J>1
\end{array}\right.
\end{equation}
\begin{equation}\label{eq5}
	Var(F) = \frac{\sum_{i \in N} F_i^2}{N} - \left( \frac{\sum_{i \in N} F_i}{N} \right) ^2.
\end{equation}
We utilize the mean and the variance of the SCFs to develop a machine-learning algorithm that can accurately identify the phase of the system. Our results highlight the remarkable effectiveness of these two parameters. 

\section*{Result}
\subsection*{Simulation}
In this section, we will clarify the inspiration behind delineating distinct phases of the systems. Assuming there are three phases, we simulate a set of systems across three different regions to identify indicators of paramagnetic, ferromagnetic, and spin-glass behaviors. We consider the SK model and simulate a fully connected pairwise interacting system of spins with quenched random couplings and $J=1$. We use the Markov chain Monte Carlo method (MCMC) to update the spins of the system at each temperature to relax the system, we are not sure that the system found the global minimum of the free energy, but we run our program in $\lceil \sqrt(N^3) \rceil$ steps. We simulated the systems in different temperatures and different $J_{0}$ and let the spins update, by using the Boltzmann factor, then calculated the mean and variance of SCFs using Eq.(\ref{eq4},\ref{eq5}). The mean and variance of the SCFs can effectively be the features to detect different phases. By recognizing the behavior of the systems within specified limits compared to the quantity of $J$, which illustrates the ratio of positive to negative interactions, we simulate different groups to evaluate how our algorithms can accurately cluster them according to the parameters we have proposed. All these simulations and clustering of systems have been carried out based on the information we already had from thermodynamics, and we knew that the system has three phases, and Our insights into the different phases stem from our understanding of the Ising model.

\begin{figure}[t]
	\centering
	\includegraphics[scale = 0.385]{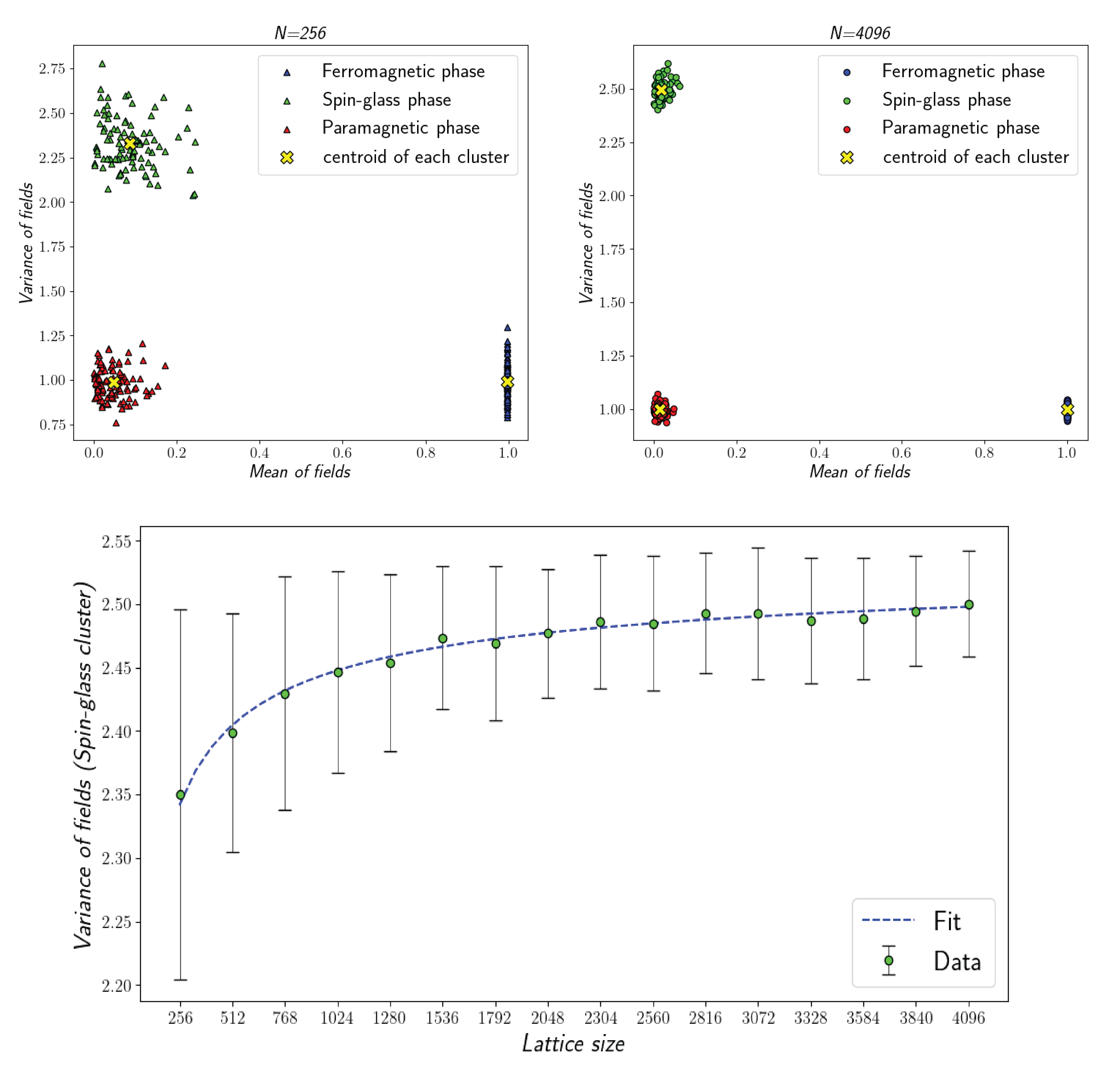}
	\caption{In the two diagrams above, red, green, and blue clusters represent the paramagnetic, spin-glass, and ferromagnetic phases, respectively. The centroid of each cluster is illustrated by \textbf{x} sign. These diagrams result in the clustering for two different lattice sizes $N=\{256, 4096\}$. The diagram below presents variations of the variance of SCFs in the spin-glass phase ($J_{0}\approx0$, $T\approx0$) as a function of lattice size. the model $y = ax^{b}+c$ fitted to data.} 
	\label{fig:fig2}
\end{figure}

When the mean value of interactions  $J_{0}$ is equal to zero, the Gaussian distribution of interactions exhibits symmetry. By increasing the mean value  $J_{0}$, the count of positive values rises; in the case of a significant increase, all interactions will become positive. This behavior is similar to the Ising model, where we know that at low temperatures, the system transitions into the ferromagnetic phase to minimize energy. Therefore, a large positive value of  $J_{0}$ in conjunction with low temperatures suggests the emergence of the ferromagnetic phase. To simulate a system in this region, we construct a network of $N$ spins and update the spins of the system at low temperatures, ideally close to zero, while $J_0$ is significantly different from zero, a large positive number. The ferromagnetic cluster phase simulated within the range $J_0 \in [1000,1000.01)$ and the temperature is close to zero $T \in (0,0.01)$. After $\lceil \sqrt(N^3) \rceil$ steps, we calculated the mean and variance of SCFs. We Repeated the simulation in a loop of $100$ iteration in the range we considered for $J_0$ and $T$ to see the cluster Fig.(\ref{fig:fig2}).
	
By raising the temperature, the system seeks a configuration that maximizes entropy, favoring the configuration with the highest entropy. In this context, the system exhibits a paramagnetic phase. In this case, we simulate a network of spins at high temperatures while $J_0$ is close to zero. The limited region that we are simulating the extreme limit of paramagnetic systems is done within the range $J_0 \in (0,0.01)$ and $T \in [1000,1000.01) $. After $\lceil \sqrt(N^3) \rceil$ updates, we calculated the mean and variance of the SCFs, and the paramagnetic clustering Fig.(\ref{fig:fig2}) results after repeating the simulation $100$ times in considered range for $J_0$ and $T$.

When $J_{0}\approx0$ and the temperature is low, a new phase known as spin-glass emerges. In this phase, the system tends to display the configuration in which opposite spins connect through negative interactions, while spins that align in the same direction tend to be connected via positive interactions. Consequently, an partial bipolar configuration can be established. It is not a complete bipolar configuration such as a model in which interaction can update\cite{Amir}, but the effect of this partial bipolar configuration can be represented in the variance of the SCFs. We simulate a group of systems running within the range $J_0 \in (0,0.01)$ and $T \in (0,0.01) $. We update the spins of the system at low temperatures, ideally close to zero, while $J_0$ is close to zero. After $\lceil \sqrt(N^3) \rceil$ updates, we calculated the mean and variance of the SCFs experienced by each spin, after $100$ iterations the spin-glass cluster appears Fig.(\ref{fig:fig2}).

In Fig.(\ref{fig:fig2}), the variance of the fields in the spin-glass phase is plotted according to the lattice size to see the size effect on the centroid and scattering of the data. We used two mathematical models, $y = a^{\prime} e^{b^{\prime}x}+c^{\prime}$, and $y = a x^b+c$ to guess the convergence of variance of the SCF in spin glass phase. The result was $a^{\prime}=-0.186$, $b^{\prime}=-0.001$, and $c^{\prime}=2.496$ for the exponential model and $a=-3.957$, $b=-0.538$, $c=2.543$ for power model. It appears that the centroid tends to converge towards a constant value, with the exponential function indicating a convergence of $2.496\pm0.358$ and the power model predicting $2.543\pm0.084$ for large lattices, which the power model is more accurate as a result of a smaller error. Because the distance of each point to the centroid of each cluster is measured relatively, the size effect can only increase the accuracy.

\begin{figure}[t]
	\centering
	\includegraphics[scale = 0.6]{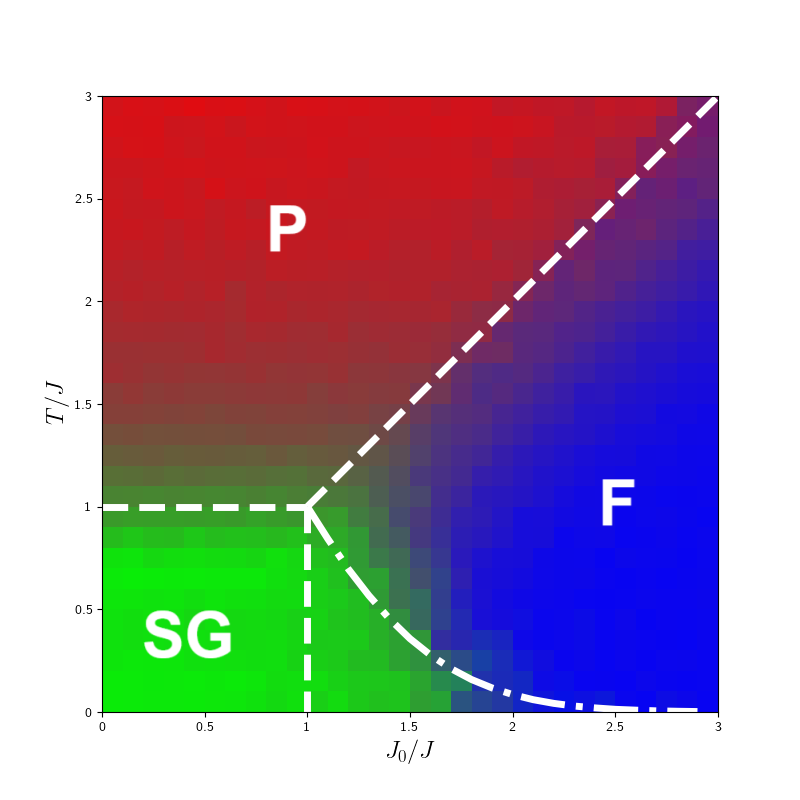}
	\caption{Phase diagram: dash lines indicate the boundaries of phase transition, which was determined by theory \cite{Sherri, SK1, PARISI1979, Parisi4, Parisi3}, and the color of each pixel shows the degree of similarity to each three phases. Red points indicate similarity to the paramagnetic phase, green points to the spin-glass phase, and blue points to the ferromagnetic phase. The diagram is $30*30$ pixels simulated for the lattice size N=1024.}
	\label{fig:fig3}
\end{figure}

  Now, we simulate systems for different $T$ and $J_{0}$ and calculate the mean and the variance of the SCFs in the system to calculate the system's distance with the centroid of the existing clusters. We use the Euclidean measure Eq.(\ref{eq6}) to calculate the distance between a newly simulated system and the centroid of other clusters. Closer to each centroid, more similar to that cluster. We use Eq.(\ref{eq7}) to calculate the similarity to each distance in which the $\epsilon$ is a small amount compared to $J$, which is 0.0001 in our simulation. The resulting diagram is the phase diagram Fig.(\ref{fig:fig3}). The phase diagram has been depicted for different $dT=0.2$ and $dJ=0.2$, and three colors indicate it. Note that each Pixel is a simulated spin-glass system with specific $T$ and $J_{0}$, and the color of each Pixel is determined by the similarity percentage to each of the three clusters. We Repeat the simulation in a loop of 20 iterations for each system to ensure accuracy and remove fluctuations. In Fig.(\ref{fig:fig3}), the dashed white line comes from the exact theoretical solution, and the dash-dot line shows the boundary between the mixed phase and the spin-glass phase. In this research, we recognize colored pixels in Fig.(\ref{fig:fig3}) follow the same behavior of the theoretical solutions in phase clustering.
\begin{equation}\label{eq6}
\text { Distance }=\sqrt{\left(X-X_c\right)^2+\left(Y-Y_c\right)^2}
\end{equation}
\begin{equation}\label{eq7}
\text { Similarity }=\frac{\epsilon}{\epsilon+\text { Distance }}
\end{equation}
The novelty of our work refers to the single replica phase detection. It provides us the best chance to give a configuration and figure out its similarity to three well-known phases. Unlike other studies that address the system's dynamic in this work, our focus is dedicated to morphology. In this method, we don't need to spend a long time letting the system drop into the global minimum of the free energy to determine the phase of the system. 

By using the mean and the variance of these SCFs as two features for clustering, we take advantage of the K-means clustering approach. We define a parameter that measures the distance of the point from each centroid concerning well-known phases and use this parameter as a similarity measure, which is valuable for finding the similarity of the Mixed phase to other phases.

\section*{Discussion}
Our study focuses on the collective behavior of a well-known complex system, spin-glass. We examine a disordered pairwise interacting system of spins. Each spin acquires information through interactions with others. To capture the various phases this system experiences, we propose a field whose mean and variance provide insights into phase detection under the interaction of a spin by others. We know that the free energy of the spin-glass system below the critical temperature has more than one minimum, which confirms the ergodicity breaking, so the system searches for energy minima. Our primary goal is to understand the phase of the system without waiting for lengthy iterations to reach a replica symmetry-breaking state.
Our findings show that the SCF can indicate the system phase. We trained a machine based on the mean and variance of SCF to distinguish these various phases. Surprisingly, in great agreement with the founded phases of a spin-glass system via analytical solution and the boundaries of phase transitions, our simulation detected similar boundaries of phase transitions. 
In a ferromagnetic state, spins feel a significant constraint when setting their direction. Consequently, the variance of the field felt by each spin is $J$, but the mean value of the fields due to the alignment of the spins is larger than other phases. The paramagnetic phase does not restrict spin alignment, and spins are randomly directed so each spin feels a near-zero mean value field from the rest; however, according to the random direction of spins, the variance of the field felt by each spin is close to $J$. For spin-glass, as we know from the phenomenology of the spin-glass, spins feel a partial force of alignment by others; therefore spins freeze in this phase, which results in a field with near zero mean value and larger variance compared to the paramagnetic and ferromagnetic state. Thus, by analyzing the mean and variance of this field, we gain insights into the distinct phases of the system.

By analyzing SCFs, we observe in Fig.(\ref{fig:fig2}) that each phase represents a distinct cluster. The ferromagnetic cluster moves to the right along the axis of the average field, which is predictable as $J_0$ increases. An interesting result is that the variance of the SCF in the spin-glass cluster is greater than that in the paramagnetic, consistent with our assumptions. We can observe the degree of similarity of the mixed phase lies between the ferromagnetic phase and the spin-glass phase. Fig.(\ref{fig:fig3}) clearly illustrates the phase diagram of the SK model, representing a phase transition from the spin-glass phase to the paramagnetic phase close to $T=J$, and a re-entrant transition from the spin-glass phase to the ferromagnetic phase close to $J_0=J$. However, while this discussion provides a comprehensive overview, there are several open questions for further exploration:
\begin{itemize}
  \item  We have found a new method to detect the phase
  diagram of the SK model with a machine-learning
  approach. The probability distribution function of
  interactions for the SK model is Gaussian. Does this
  method work for finding the phase diagram of the
  other spin-glass models with different probability
  distribution functions of interactions?
  \item Does the variance of the SCF change when the interactions are correlated? 
  \item Can this method be used to determine the phase diagram of the P-spin model?
  \item Can we use the social, biological, and economic data to predict the phase of these systems, and define an early warning parameter for phase transition?\\
\end{itemize}

\bibliography{MyReferences}

\begin{thebibliography}{10}
\urlstyle{rm}
\expandafter\ifx\csname url\endcsname\relax
  \def\url#1{\texttt{#1}}\fi
\expandafter\ifx\csname urlprefix\endcsname\relax\def\urlprefix{URL }\fi
\expandafter\ifx\csname doiprefix\endcsname\relax\def\doiprefix{DOI: }\fi
\providecommand{\bibinfo}[2]{#2}
\providecommand{\eprint}[2][]{\url{#2}}

\bibitem{AndersonEdwards}
\bibinfo{author}{Edwards, S.~F.} \& \bibinfo{author}{Anderson, P.~W.}
\newblock \bibinfo{journal}{\bibinfo{title}{Theory of spin glasses}}.
\newblock {\emph{\JournalTitle{Journal of Physics F: Metal Physics}}} \textbf{\bibinfo{volume}{5}}, \bibinfo{pages}{965}, \doiprefix\url{https://doi.org/10.1088/0305-4608/5/5/017} (\bibinfo{year}{1975}).

\bibitem{SK1}
\bibinfo{author}{Sherrington, D.} \& \bibinfo{author}{Kirkpatrick, S.}
\newblock \bibinfo{journal}{\bibinfo{title}{Solvable model of a spin-glass}}.
\newblock {\emph{\JournalTitle{Phys. Rev. Lett.}}} \textbf{\bibinfo{volume}{35}}, \bibinfo{pages}{1792--1796}, \doiprefix\url{https://doi.org/10.1103/PhysRevLett.35.1792} (\bibinfo{year}{1975}).

\bibitem{Sherri}
\bibinfo{author}{Kirkpatrick, S.} \& \bibinfo{author}{Sherrington, D.}
\newblock \bibinfo{journal}{\bibinfo{title}{Infinite-ranged models of spin-glasses}}.
\newblock {\emph{\JournalTitle{Phys. Rev. B}}} \textbf{\bibinfo{volume}{17}}, \bibinfo{pages}{4384--4403}, \doiprefix\url{https://doi.org/10.1103/PhysRevB.17.4384} (\bibinfo{year}{1978}).

\bibitem{ParisiReplica}
\bibinfo{author}{Parisi, G.}
\newblock \bibinfo{journal}{\bibinfo{title}{On the replica approach to spin-glass theory}}.
\newblock {\emph{\JournalTitle{Philosophical Magazine B}}} \textbf{\bibinfo{volume}{71}}, \bibinfo{pages}{471--478}, \doiprefix\url{https://doi.org/10.1080/01418639508238538} (\bibinfo{year}{1995}).

\bibitem{PARISI1979}
\bibinfo{author}{Parisi, G.}
\newblock \bibinfo{journal}{\bibinfo{title}{Toward a mean field theory for spin glasses}}.
\newblock {\emph{\JournalTitle{Physics Letters A}}} \textbf{\bibinfo{volume}{73}}, \bibinfo{pages}{203--205}, \doiprefix\url{https://doi.org/10.1016/0375-9601(79)90708-4} (\bibinfo{year}{1979}).

\bibitem{VanHemmen}
\bibinfo{author}{van Hemmen, J.~L.} \& \bibinfo{author}{Palmer, R.~G.}
\newblock \bibinfo{journal}{\bibinfo{title}{The replica method and solvable spin glass model}}.
\newblock {\emph{\JournalTitle{Journal of Physics A: Mathematical and General}}} \textbf{\bibinfo{volume}{12}}, \bibinfo{pages}{563}, \doiprefix\url{https://doi.org/10.1088/0305-4470/12/4/016} (\bibinfo{year}{1979}).

\bibitem{Parisi1}
\bibinfo{author}{M\'ezard, M.}, \bibinfo{author}{Parisi, G.}, \bibinfo{author}{Sourlas, N.}, \bibinfo{author}{Toulouse, G.} \& \bibinfo{author}{Virasoro, M.}
\newblock \bibinfo{journal}{\bibinfo{title}{Nature of the spin-glass phase}}.
\newblock {\emph{\JournalTitle{Phys. Rev. Lett.}}} \textbf{\bibinfo{volume}{52}}, \bibinfo{pages}{1156--1159}, \doiprefix\url{https://doi.org/10.1103/PhysRevLett.52.1156} (\bibinfo{year}{1984}).

\bibitem{Binder1986}
\bibinfo{author}{Binder, K.} \& \bibinfo{author}{Young, A.~P.}
\newblock \bibinfo{journal}{\bibinfo{title}{Spin glasses: Experimental facts, theoretical concepts, and open questions}}.
\newblock {\emph{\JournalTitle{Rev. Mod. Phys.}}} \textbf{\bibinfo{volume}{58}}, \bibinfo{pages}{801--976}, \doiprefix\url{https://doi.org/10.1103/RevModPhys.58.801} (\bibinfo{year}{1986}).

\bibitem{TAP}
\bibinfo{author}{Thouless, D.~J.}, \bibinfo{author}{Anderson, P.~W.} \& \bibinfo{author}{Palmer, R.~G.}
\newblock \bibinfo{journal}{\bibinfo{title}{Solution of 'solvable model of a spin glass'}}.
\newblock {\emph{\JournalTitle{The Philosophical Magazine: A Journal of Theoretical Experimental and Applied Physics}}} \textbf{\bibinfo{volume}{35}}, \bibinfo{pages}{593--601}, \doiprefix\url{https://doi.org/10.1080/14786437708235992} (\bibinfo{year}{1977}).

\bibitem{cavity}
\bibinfo{author}{Mézard, M.}, \bibinfo{author}{Parisi, G.} \& \bibinfo{author}{Virasoro, M.~A.}
\newblock \bibinfo{journal}{\bibinfo{title}{Sk model: The replica solution without replicas}}.
\newblock {\emph{\JournalTitle{Europhysics Letters}}} \textbf{\bibinfo{volume}{1}}, \bibinfo{pages}{77}, \doiprefix\url{https://doi.org/10.1209/0295-5075/1/2/006} (\bibinfo{year}{1986}).

\bibitem{Spin_Glass_Theory_and_Beyond}
\bibinfo{author}{Mezard, M.}, \bibinfo{author}{Parisi, G.} \& \bibinfo{author}{Virasoro, M.}
\newblock \emph{\bibinfo{title}{Spin Glass Theory and Beyond}} (\bibinfo{publisher}{WORLD SCIENTIFIC}, \bibinfo{year}{1986}).
\newblock \eprint{https://www.worldscientific.com/doi/pdf/10.1142/0271}.

\bibitem{Nishimori}
\bibinfo{author}{Nishimori, H.}
\newblock \emph{\bibinfo{title}{Statistical Physics of Spin Glasses and Information Processing: An Introduction}} (\bibinfo{publisher}{Oxford University Press}, \bibinfo{year}{2001}).

\bibitem{Parisi4}
\bibinfo{author}{Parisi, G.}
\newblock \bibinfo{journal}{\bibinfo{title}{Order parameter for spin-glasses}}.
\newblock {\emph{\JournalTitle{Phys. Rev. Lett.}}} \textbf{\bibinfo{volume}{50}}, \bibinfo{pages}{1946--1948}, \doiprefix\url{https://doi.org/10.1103/PhysRevLett.50.1946} (\bibinfo{year}{1983}).

\bibitem{Parisi3}
\bibinfo{author}{Parisi, G.}
\newblock \bibinfo{journal}{\bibinfo{title}{The order parameter for spin glasses: a function on the interval 0-1}}.
\newblock {\emph{\JournalTitle{Journal of Physics A: Mathematical and General}}} \textbf{\bibinfo{volume}{13}}, \bibinfo{pages}{1101}, \doiprefix\url{https://doi.org/10.1088/0305-4470/13/3/042} (\bibinfo{year}{1980}).

\bibitem{Parisi1980}
\bibinfo{author}{{Parisi}, G.}
\newblock \bibinfo{journal}{\bibinfo{title}{{A sequence of approximated solutions to the S-K model for spin glasses}}}.
\newblock {\emph{\JournalTitle{Journal of Physics A Mathematical General}}} \textbf{\bibinfo{volume}{13}}, \bibinfo{pages}{L115--L121}, \doiprefix\url{https://doi.org/10.1088/0305-4470/13/4/009} (\bibinfo{year}{1980}).

\bibitem{PhysRevLettTam}
\bibinfo{author}{Tam, K.-M.} \& \bibinfo{author}{Gingras, M. J.~P.}
\newblock \bibinfo{journal}{\bibinfo{title}{Spin-glass transition at nonzero temperature in a disordered dipolar ising system: The case of ${\mathrm{liho}}_{x}{\mathbf{y}}_{1\ensuremath{-}x}{\mathbf{f}}_{4}$}}.
\newblock {\emph{\JournalTitle{Phys. Rev. Lett.}}} \textbf{\bibinfo{volume}{103}}, \bibinfo{pages}{087202}, \doiprefix\url{https://doi.org/10.1103/PhysRevLett.103.087202} (\bibinfo{year}{2009}).

\bibitem{Drozd}
\bibinfo{author}{Drozd-Rzoska, A.}, \bibinfo{author}{Rzoska, S.~J.}, \bibinfo{author}{Pawlus, S.}, \bibinfo{author}{Martinez-Garcia, J.~C.} \& \bibinfo{author}{Tamarit, J.~L.}
\newblock \bibinfo{journal}{\bibinfo{title}{Evidence for critical-like behavior in ultraslowing glass-forming systems.}}
\newblock {\emph{\JournalTitle{Physical review. E, Statistical, nonlinear, and soft matter physics}}} \textbf{\bibinfo{volume}{82 3 Pt 1}}, \bibinfo{pages}{031501}, \doiprefix\url{https://doi.org/10.1103/PhysRevE.82.031501} (\bibinfo{year}{2010}).

\bibitem{park2022observation}
\bibinfo{author}{Park, J.-H.} \emph{et~al.}
\newblock \bibinfo{journal}{\bibinfo{title}{Observation of spin-glass-like characteristics in ferrimagnetic tbco through energy-level-selective approach}}.
\newblock {\emph{\JournalTitle{Nature Communications}}} \textbf{\bibinfo{volume}{13}}, \bibinfo{pages}{5530}, \doiprefix\url{https://doi.org/10.1038/s41467-022-33195-y} (\bibinfo{year}{2022}).

\bibitem{Mydosh2015}
\bibinfo{author}{Mydosh, J.~A.}
\newblock \bibinfo{journal}{\bibinfo{title}{Spin glasses: redux: an updated experimental/materials survey}}.
\newblock {\emph{\JournalTitle{Reports on Progress in Physics}}} \textbf{\bibinfo{volume}{78}}, \bibinfo{pages}{052501}, \doiprefix\url{https://doi.org/10.1088/0034-4885/78/5/052501} (\bibinfo{year}{2015}).

\bibitem{katukuri2015strong}
\bibinfo{author}{Katukuri, V.~M.} \emph{et~al.}
\newblock \bibinfo{journal}{\bibinfo{title}{Strong magnetic frustration and anti-site disorder causing spin-glass behavior in honeycomb li2rho3}}.
\newblock {\emph{\JournalTitle{Scientific reports}}} \textbf{\bibinfo{volume}{5}}, \bibinfo{pages}{14718}, \doiprefix\url{https://doi.org/10.1038/srep14718} (\bibinfo{year}{2015}).

\bibitem{Hudetz2014}
\bibinfo{author}{Hudetz, A.}, \bibinfo{author}{Humphries, C.~J.} \& \bibinfo{author}{Binder, J.}
\newblock \bibinfo{journal}{\bibinfo{title}{Spin-glass model predicts metastable brain states that diminish in anesthesia}}.
\newblock {\emph{\JournalTitle{Frontiers in Systems Neuroscience}}} \doiprefix\url{10.3389/fnsys.2014.00234} (\bibinfo{year}{2014}).

\bibitem{knuuttila2014}
\bibinfo{author}{Knuuttila, T.} \& \bibinfo{author}{Loettgers, A.}
\newblock \bibinfo{journal}{\bibinfo{title}{Magnets, spins, and neurons: The dissemination of model templates across disciplines}}.
\newblock {\emph{\JournalTitle{The Monist}}} \textbf{\bibinfo{volume}{97}}, \bibinfo{pages}{280--300} (\bibinfo{year}{2014}).

\bibitem{Lidar}
\bibinfo{author}{Lidar, D.~A.} \& \bibinfo{author}{Biham, O.}
\newblock \bibinfo{journal}{\bibinfo{title}{Simulating ising spin glasses on a quantum computer}}.
\newblock {\emph{\JournalTitle{Phys. Rev. E}}} \textbf{\bibinfo{volume}{56}}, \bibinfo{pages}{3661--3681}, \doiprefix\url{https://doi.org/10.1103/PhysRevE.56.3661} (\bibinfo{year}{1997}).

\bibitem{Callison2019}
\bibinfo{author}{Callison, A.}, \bibinfo{author}{Chancellor, N.}, \bibinfo{author}{Mintert, F.} \& \bibinfo{author}{Kendon, V.}
\newblock \bibinfo{journal}{\bibinfo{title}{Finding spin glass ground states using quantum walks}}.
\newblock {\emph{\JournalTitle{New Journal of Physics}}} \textbf{\bibinfo{volume}{21}}, \bibinfo{pages}{123022}, \doiprefix\url{https://doi.org/10.1088/1367-2630/ab5ca2} (\bibinfo{year}{2019}).

\bibitem{kadowaki1998quantum}
\bibinfo{author}{Kadowaki, T.} \& \bibinfo{author}{Nishimori, H.}
\newblock \bibinfo{journal}{\bibinfo{title}{Quantum annealing in the transverse ising model}}.
\newblock {\emph{\JournalTitle{Physical Review E}}} \textbf{\bibinfo{volume}{58}}, \bibinfo{pages}{5355}, \doiprefix\url{https://doi.org/10.1103/PhysRevE.58.5355} (\bibinfo{year}{1998}).

\bibitem{king2023quantum}
\bibinfo{author}{King, A.~D.} \emph{et~al.}
\newblock \bibinfo{journal}{\bibinfo{title}{Quantum critical dynamics in a 5,000-qubit programmable spin glass}}.
\newblock {\emph{\JournalTitle{Nature}}} \textbf{\bibinfo{volume}{617}}, \bibinfo{pages}{61--66}, \doiprefix\url{https://doi.org/10.1038/s41586-023-05867-2} (\bibinfo{year}{2023}).

\bibitem{harris2018phase}
\bibinfo{author}{Harris, R.} \emph{et~al.}
\newblock \bibinfo{journal}{\bibinfo{title}{Phase transitions in a programmable quantum spin glass simulator}}.
\newblock {\emph{\JournalTitle{Science}}} \textbf{\bibinfo{volume}{361}}, \bibinfo{pages}{162--165}, \doiprefix\url{https://doi.org/10.1126/science.aat2025} (\bibinfo{year}{2018}).

\bibitem{knysh2016zero}
\bibinfo{author}{Knysh, S.}
\newblock \bibinfo{journal}{\bibinfo{title}{Zero-temperature quantum annealing bottlenecks in the spin-glass phase}}.
\newblock {\emph{\JournalTitle{Nature communications}}} \textbf{\bibinfo{volume}{7}}, \bibinfo{pages}{12370}, \doiprefix\url{https://doi.org/10.1038/ncomms12370} (\bibinfo{year}{2016}).

\bibitem{grass2016quantum}
\bibinfo{author}{Gra{\ss}, T.}, \bibinfo{author}{Ravent{\'o}s, D.}, \bibinfo{author}{Juli{\'a}-D{\'\i}az, B.}, \bibinfo{author}{Gogolin, C.} \& \bibinfo{author}{Lewenstein, M.}
\newblock \bibinfo{journal}{\bibinfo{title}{Quantum annealing for the number-partitioning problem using a tunable spin glass of ions}}.
\newblock {\emph{\JournalTitle{Nature communications}}} \textbf{\bibinfo{volume}{7}}, \bibinfo{pages}{11524}, \doiprefix\url{https://doi.org/10.1038/ncomms11524} (\bibinfo{year}{2016}).

\bibitem{GROSS}
\bibinfo{author}{Gross, D.} \& \bibinfo{author}{Mezard, M.}
\newblock \bibinfo{journal}{\bibinfo{title}{The simplest spin glass}}.
\newblock {\emph{\JournalTitle{Nuclear Physics B}}} \textbf{\bibinfo{volume}{240}}, \bibinfo{pages}{431--452}, \doiprefix\url{https://doi.org/10.1016/0550-3213(84)90237-2} (\bibinfo{year}{1984}).

\bibitem{GARDNER}
\bibinfo{author}{Gardner, E.}
\newblock \bibinfo{journal}{\bibinfo{title}{Spin glasses with p-spin interactions}}.
\newblock {\emph{\JournalTitle{Nuclear Physics B}}} \textbf{\bibinfo{volume}{257}}, \bibinfo{pages}{747--765}, \doiprefix\url{https://doi.org/10.1016/0550-3213(85)90374-8} (\bibinfo{year}{1985}).

\bibitem{Crisanti}
\bibinfo{author}{Crisanti, A.} \& \bibinfo{author}{Sommers, H.-J.}
\newblock \bibinfo{journal}{\bibinfo{title}{The sphericalp-spin interaction spin glass model: the statics}}.
\newblock {\emph{\JournalTitle{Zeitschrift f{\"u}r Physik B Condensed Matter}}} \textbf{\bibinfo{volume}{87}}, \bibinfo{pages}{341--354}, \doiprefix\url{https://doi.org/10.1007/BF01309287} (\bibinfo{year}{1992}).

\bibitem{Mahsa2}
\bibinfo{author}{Bagherikalhor, M.}, \bibinfo{author}{Askari, B.} \& \bibinfo{author}{Jafari, G.~R.}
\newblock \bibinfo{journal}{\bibinfo{title}{Triadic interaction in the background of a pairwise spin-glass}}.
\newblock {\emph{\JournalTitle{Phys. Rev. E}}} \textbf{\bibinfo{volume}{109}}, \bibinfo{pages}{064105}, \doiprefix\url{https://doi.org/10.1103/PhysRevE.109.064105} (\bibinfo{year}{2024}).

\bibitem{Thouless_1978}
\bibinfo{author}{de~Almeida, J. R.~L.} \& \bibinfo{author}{Thouless, D.~J.}
\newblock \bibinfo{journal}{\bibinfo{title}{Stability of the sherrington-kirkpatrick solution of a spin glass model}}.
\newblock {\emph{\JournalTitle{Journal of Physics A: Mathematical and General}}} \textbf{\bibinfo{volume}{11}}, \bibinfo{pages}{983}, \doiprefix\url{https://doi.org/10.1088/0305-4470/11/5/028} (\bibinfo{year}{1978}).

\bibitem{Amir}
\bibinfo{author}{Kargaran, A.} \& \bibinfo{author}{Jafari, G.~R.}
\newblock \bibinfo{journal}{\bibinfo{title}{Heider and coevolutionary balance: From discrete to continuous phase transition}}.
\newblock {\emph{\JournalTitle{Phys. Rev. E}}} \textbf{\bibinfo{volume}{103}}, \bibinfo{pages}{052302}, \doiprefix\url{https://doi.org/10.1103/PhysRevE.103.052302} (\bibinfo{year}{2021}).

\end{thebibliography}

%\noindent LaTeX formats citations and references automatically using the bibliography records in your .bib file, which you can edit via the project menu. Use the cite command for an inline citation, e.g.  \cite{Hao:gidmaps:2014}.

%For data citations of datasets uploaded to e.g. \emph{figshare}, please use the \verb|howpublished| option in the bib entry to specify the platform and the link, as in the \verb|Hao:gidmaps:2014| example in the sample bibliography file.

%\section*{Acknowledgements (not compulsory)}

%Acknowledgements should be brief, and should not include thanks to anonymous referees and editors, or effusive comments. Grant or contribution numbers may be acknowledged.

\section*{Author contributions statement}
Conceptualization: Ali Talebi, Mahsa Bagerikalhor, Behrouz Asgari, Reza Jafari. 
Formal analysis: Ali Talebi. 
Investigation: Ali Talebi, Mahsa Bagerikalhor, Behrouz Asgari, Reza Jafari. 
Project administration: Reza Jafari. 
Visualization: Ali Talebi 
Writing – original draft: Ali Talebi, Mahsa Bagerikalhor. Writing – review \& editing: Ali Talebi, Reza Jafari.
All authors read and reviewed the manuscript. 

\section*{Additional information} 
We made an Executable File to receive the lattice size, mean of interactions, and temperature from the user to determine the similarity to each well-known phase for SK model based on the algorithm we introduced in this article. Link: \href{https://github.com/alitalebi2000/spin-glass-phase-detection-ML.git}{github.com/alitalebi2000/spin-glass-phase-detection-ML.git}
\end{document}